\documentclass[11pt,onecolumn,draftcls]{IEEEtran}

\usepackage{graphicx,amsmath}

\begin{document}

\title{Rate Compatible Protocol for Information Reconciliation: An application to QKD}

\author{
   \IEEEauthorblockN{
   David Elkouss,
    Jesus Martinez-Mateo,
    Daniel Lancho and
    Vicente Martin
  }
  \\
  \IEEEauthorblockA{
    Facultad de Inform\'atica, Universidad Polit\'ecnica de Madrid,\\
    Campus de Montegancedo, 28660 Boadilla del Monte (Madrid), Spain,\\ 
    e-mail:\{delkouss, jmartinez, dlancho, vicente\}@fi.upm.es
  }
}

\maketitle

\begin{abstract}
Information Reconciliation is a mechanism that allows to weed out the discrepancies between two correlated variables. It is an essential component in every key agreement protocol where the key has to be transmitted through a noisy channel. The typical case is in the satellite scenario described by Maurer in the early 90's. Recently the need has arisen in relation with Quantum Key Distribution (QKD) protocols, where it is very important not to reveal unnecessary information in order to maximize the shared key length. In this paper we present an information reconciliation protocol based on a rate compatible construction of Low Density Parity Check codes. Our protocol improves the efficiency of the reconciliation for the whole range of error rates in the discrete variable QKD context. Its adaptability together with its low interactivity makes it specially well suited for QKD reconciliation.
\end{abstract}

\begin{IEEEkeywords}
Reconciliation, low-density parity-check (LDPC) codes, puncturing, shortening, rate-compatible.
\end{IEEEkeywords}

\IEEEpeerreviewmaketitle

\section{Introduction}

The general scenario for information reconciliation is one in which two parties have two sets of correlated data with some discrepancies between them. The situation is equivalent to transmit the data from one party to the other through a noisy channel, akin in the satellite scenario described by Maurer~\cite{Maurer_93}.

In a Quantum Key Distribution (QKD) protocol, errors are generated in the communications channel either by the interaction of the quantum information carrier with the environment, by imperfections in the QKD device or by an eavesdropper. The two parties participating in the communication, Alice and Bob, thus have two sets of correlated data from which a common set must be extracted. This problem has been previously subject to consideration \cite{Brassard_94,Watanabe_07,Watanabe_08,Leverrier_08,Elkouss_09}. It is a process known as key distillation, that requires a discussion carried over an authenticated classical channel. It is interactive in the sense that it needs communications through the channel. Since it can also be listened by an eavesdropper, it is important to minimize the amount of information that have to be transmitted in the reconciliation process. Any extra information limits the performance of the QKD implementation. In theory one could minimize the information leakage using a highly interactive protocol, but in practical applications this would lead to a prohibitively large communication overhead through the network, limiting also the effective keyrate.

It is in this scenario where modern Forward Error Correction (FEC) is an interesting solution. The idea is to make use of FEC's inherent advantage of requiring a single channel use to reconcile the two sets. 
In~\cite{Elkouss_09} it was analyzed the use of a discrete number of Low-Density Parity-Check (LDPC) codes optimized for the binary symmetric channel. As a consequence the efficiency exhibited an staircase-like behaviour: each code was used within a range of error rates and the reconciliation efficiency was maximized only in the region close to the code's threshold. 

In this work, we develop the idea of using LDPC codes optimized for the binary symmetric channel. We take these codes as an starting point and develop a rate compatible information reconciliation protocol with an efficiency close to optimal. In particular, the proposed protocol builds codes that minimize the exchanged information for error probabilities between $1\%$ and $10\%$\footnote{The maximum error thresholds for extracting an absolute secret key in a QKD protocol is $11\%$~\cite{Gisin_02}.}, the expected values in real implementations of QKD systems. 

This solution addresses the rate adaptation problem (open problem 2) from the recent review paper of Matsumoto~\cite{Matsumoto_09} in which he lists the problems that an LDPC solution should overcome in order to compare advantageously to current interactive reconciliation solutions.

The paper is organised as follows:
In Section~\ref{sec-ratecompir} the main ideas are discussed. A new Information Reconciliation Protocol able to adapt to different channel parameters is presented and its asymptotic behavior discussed.
In Section~\ref{sec-results} the results of a practical implementation of the protocol are shown. In particular we have analyzed the rate compared to the optimal value and the reconciliation efficiency.



\section{Rate Compatible Information Reconciliation}
\label{sec-ratecompir}
\subsection*{Information Reconciliation}

Let $X$ and $Y$ be two of correlated variables belonging to Alice and Bob, and $\mathbf{x}$ and $\mathbf{y}$ their outcome strings, Information Reconciliation~\cite{Brassard_94} is a mechanism that allows them to eliminate the discrepancies between $\mathbf{x}$ and $\mathbf{y}$ and agree on a string $S(\mathbf{x})$ ---with possibly $S(\mathbf{x}) = \mathbf{x}$.

\begin{figure}
\centering
\includegraphics[width=0.6\linewidth]{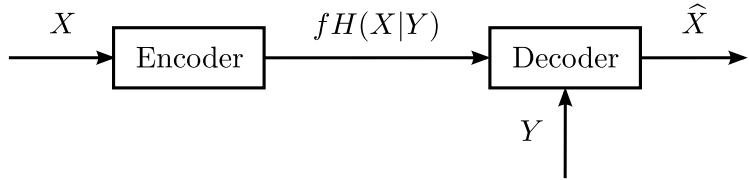}
\caption{Source coding with side information.}
\label{fig:side-information}
\end{figure}

The problem of information reconciliation can be seen as the source coding problem with side information (see Fig. ~\ref{fig:side-information}). Thus, as shown by Slepian and Wolf~\cite{Slepian_73}, the minimum information $I$ that Alice would have to send to Bob in order to help him reconcile $Y$ and $X$ is $I_{opt}=H(X|Y)$. Taking into account that real reconciliation will not be optimal, we use a parameter $f\geq 1$ as a quality figure for the reconciliation efficiency:

\begin{equation}
I_{real} = f H(X|Y) \geq I_{opt}
\label{eq:efficiency}
\end{equation}

Here we will concentrate on binary variables, which apply to discrete variable QKD, although the ideas are directly applicable to other scenarios. 

The most widely used protocol for information reconciliation in QKD is \textit{Cascade}~\cite{Brassard_94}, because of its simplicity and good efficiency.  \textit{Cascade} is a highly interactive protocol that runs for a certain number of passes. In each pass, Alice and Bob both perform the same permutation on their respective strings, divide them in blocks of the same size and exchange the parities of the blocks. Whenever there is a mismatch they perform a dichotomic search to find an error, finding one usually means discovering more errors left in previous passes.

The main handicap of \textit{Cascade} is its high interactivity. Buttler et al~\cite{Buttler_02} proposed Winnow, a reconciliation protocol where instead of exchanging block parities, Alice and Bob exchange the syndrome of a Hamming code. Their protocol succeeded in reducing the interactivity but, in the error range of interest for QKD, the efficiency was worse than that of \textit{Cascade}.

There has been further work on improving the efficiency of \textit{Cascade}-like protocols. In~\cite{Sugimoto_00} the block size is optimized, while in~\cite{Liu_03} the emphasis is put on minimizing the information sent to correct one error on each pass.

\subsection*{Definitions}

LDPC codes were introduced by Gallager in the early 60's~\cite{Gallager_63}. They are linear codes with a sparse parity check matrix.

A family of LDPC codes is defined by two \textit{generating polynomials}~\cite{Richardson_01a}, $\lambda(x)$ and $\rho(x)$:

\begin{equation}
\lambda(x) = \sum_{i = 2}^{d_{s_{max}}} \lambda_i x^{i-1}
\;
;
\quad
\rho(x) = \sum_{j = 2}^{d_{c_{max}}} \rho_j x^{j-1}
\label{eq:polynomials}
\end{equation}

\noindent where $\lambda(x)$ and $\rho(x)$ define degree distributions. $\lambda_i$ and $\rho_i$ indicate the proportion (normalized to $1$) of edges connected to symbol and check nodes of degree $i$, respectively. The rate $R_0$ of the family of LDPC codes is defined as:

\begin{equation}
R_0 = 1 - \frac{\sum_i \lambda_i / i}{\sum_j \rho_j / j}
\label{eq:rate-polynomials}
\end{equation}



\begin{figure}

\includegraphics[width=0.6\linewidth]{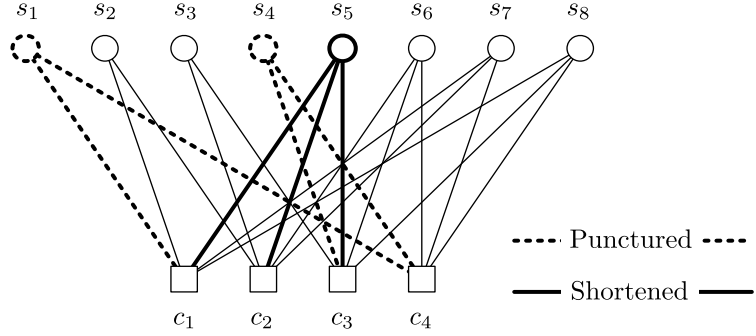}
\centering
\caption{Example of puncturing and shortening applied to a code represented by a Tanner graph. The rate of the original code is $R=(n-m)/n=(8-4)/8=1/2$. After puncturing two symbol nodes (indicated in the graph with dashed lines) the new rate is increased to $R=(8-4)/(8-2)=2/3$. Shortening one symbol of the original code (indicated with thick solid lines) leads to a new rate of $R=((8-1)-4)/(8-1)=3/7$. Puncturing two symbols and shortening one the original code leads to a rate of $R=((8-1)-4)/(8-2-1)=3/5$.}
\label{fig:tanner-graph}
\end{figure}

Two common strategies to adapt the rate to the channel parameters are puncturing and shortening~\cite{Tian_05}. Puncturing means deleting a predefined set of $p$ symbols from each word, converting a $[n,k]$ code into a $[n-p,k]$ code. Shortening means deleting a set of $s$ symbols from the encoding process, converting a $[n,k]$ code into a $[n-s,k-s]$ code. 
Both processes allow to modulate the rate of the code as:

\begin{equation}
R = \frac{R_0 - \sigma}{1 - \pi - \sigma} = \frac{k - s}{n - p - s}
\label{eq:rate}
\end{equation}

\noindent where $\pi$ and $\sigma$ represent the ratios of information punctured and shortened respectively, and $R_0$ is the rate of the initial code (see Fig.~\ref{fig:tanner-graph} for an example).

\subsection*{The protocol}

\begin{figure}
\centering
\includegraphics[width=0.6\linewidth]{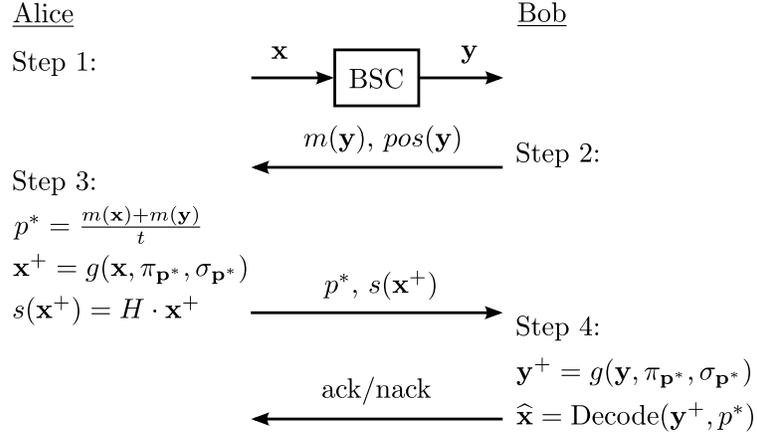}
\caption{Protocol sequence diagram.}
\label{fig:protocol}
\end{figure}

Standard puncturing and shortening need an a priori knowledge about the channel in order to adapt the rate. The Bit Error Rate (BER) in the case of QKD protocols is an a priori unknown value, hence it is important to be able to construct codes that can adapt to the varying BER values that might appear during a QKD transmission. In order to cope with this, we propose an inverse puncturing and shortening protocol, that is performed after the distribution of the correlated variables.

\begin{figure}
\centering
\includegraphics[width=0.6\linewidth]{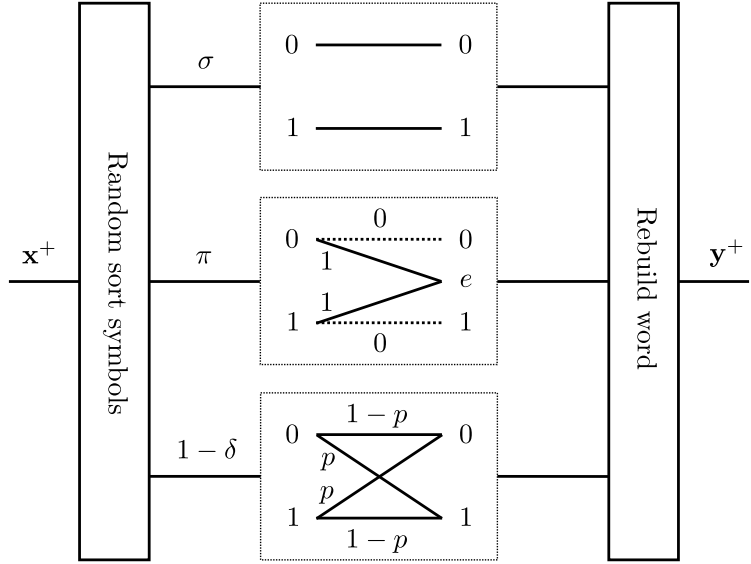}
\caption{Channel model. The protocol described can be interpreted as a communication through three channels with different probabilities: a noiseless channel with probability $\sigma$, a $\textrm{BEC}(1)$ with probability $\pi$, and a $\textrm{BSC}(p)$ with probability $1 - \delta$.}
\label{fig:coding-model}
\end{figure}

The protocol assumes the existence of a shared pool of codes of length $n$, adjusted for different rates. Depending on the range of crossover probabilities to be corrected, a parameter $\delta$ is chosen to set the proportion of bits to be either shortened ($\sigma$) or punctured ($\pi$; $\delta = \pi + \sigma$). $\delta$ defines the achievable rates, $R$, through:

\begin{equation}
\dfrac{R_0-\delta}{1-\delta}\leq R \leq \dfrac{R_0}{1-\delta}
\end{equation}

\noindent with $R_0$ being the rate of the code selected from the pool. 
 For an $[n, k]$ code this would mean $n \cdot \pi$ bits punctured, $n \cdot \sigma$ bits shortened and $n \cdot (1 - \delta)$ bits transmitted over the BSC (see Fig.~\ref{fig:coding-model}). The number of symbols not to be sent is $d=\lfloor \delta \cdot n \rfloor$.

The protocol goes through the following steps:

\subsubsection*{Step 1}

Alice sends to Bob a message $\mathbf{x}$, an instance of variable $X$, of size $\ell = n - d$ through a BSC of crossover probability $p$ (or a black box behaving as such). Bob receives the correlated message, $\mathbf{y}$.

\subsubsection*{Step 2}

Bob chooses randomly $t$ bits of $\mathbf{y}$, $m(\mathbf{y})$, and sends them and their positions, $pos(\mathbf{y})$, to Alice. 

\subsubsection*{Step 3}

Using $pos(\mathbf{y})$, Alice extracts $m(\mathbf{x})$ and estimates the crossover probability: 

\begin{equation}
p^* = \dfrac{m(\mathbf{x}) + m(\mathbf{y})}{t}
\end{equation}

Once Alice has estimated $p^*$, she knows the theoretical rate for a punctured and shortened code able to correct the string. Now she must decide what is the optimal rate corresponding to the efficiency of the code she is using: $R = 1 - f(p^*) h(p^*)$; where $h$ is the binary entropy function and $f$ the efficiency (e.g. Tab.~\ref{tab:efficiency}). Then she can derive the optimal values for $s$ and $p$:

\begin{equation}
\begin{split}
s & = \lceil (R_0 - R (1 - d / n)) \cdot n \rceil \\
p & = d - s
\end{split}
\end{equation}

Alice creates now a string $\mathbf{x^+}=g(\mathbf{x},\mathbf{\sigma_{p^*}},\mathbf{\pi_{p^*}})$ of size $n$. The function $g$ defines the $n-d$ positions are going to have the values of string $\mathbf{x}$, the $p$ positions that are going to be assigned random values, and the $s$ positions that are going to have values known by Alice and Bob. The set of $n-d$ positions, the set of $p$ positions and the set of $s$ positions and their values come from a synchronized pseudo-random generator. She then sends $s(\mathbf{x^+})$, the syndrome of $\mathbf{x^+}$, to Bob as well as the estimated crossover probability $p^*$.


\subsubsection*{Step 4}

Bob can reproduce Alice's estimation of the optimal rate $R$, the positions of the $p$ punctured bits, and the positions and values of the $s$ shortened bits, and then he creates the corresponding string $\mathbf{y^+} = g(\mathbf{y}, \mathbf{\sigma_{p^*}}, \mathbf{\pi_{p^*}})$.

Bob should now be able to decode Alice's codeword with high probability, as the rate has been adapted to the channel crossover probability. He finally sends an acknowledge to Alice to indicate if he successfully recovered $\mathbf{x^+}$.

\subsubsection*{Example}


Calculation of $s$ and $p$ for step 3.
Alice and Bob use a $[10^6, 5 \times 10^5]$ code, $d=10^5$, and they have found out that the efficiency of their reconciliation behaves as $f(p)=1.1 + |p - 0.1|$. When Alice estimates the discrepancy, she finds that $p^*=0.08$. If the code were optimal, it would have been designed with a rate $R = 1 - f(0.08) h(0.08) = 1 - (1.12)(0.402) = 0.55$. Then she obtains $s = 2.25 \times 10^5$, and $p = 2.75 \times 10^5$.


In the case in which the protocol is used to reconcile secret keys, several modifications have to be done. In step 1 the size should be increased by $t$, $\ell = n - d + t$. In step 2, Bob should discard from his string, $\mathbf{x}$, the $t$ bits that have been published. Finally, in step 3, Alice should also discard the $t$ published bits from hers.



\subsection*{Performance analysis}

We are first interested in the range of rates in which the protocol can be used and the expected efficiency if the codes are long enough. 
The threshold value is calculated using the density evolution algorithm~\cite{Richardson_01a}, and in particular we have implemented the discretized version of Chung et al~\cite{Chung_01}. The equation used to track the evolution of the density function is:

\begin{equation}
p_u^{(l + 1)} = \rho (p_{u_0} \ast \lambda (p_u^{(l)}))
\label{eq:discretized-density-evolution}
\end{equation}

\noindent where $p_u^{(l)}$ is the probability mass function at the symbols during iteration $l$, and $p_{u_0}$ is the initial message density distribution, which in our case is: 

\begin{equation}
p_{u_0}(x) = (1 - \delta) p_{u_0}^{\textrm{BSC}}(x) + \pi \Delta_{0}(x) + \sigma \Delta_{\infty}(x)
\label{eq:p0_ps}
\end{equation}

\noindent where $p_{u_0}^{\textrm{BSC}}(x) = p \Delta_{\mathrm{-log}\frac{p}{1-p}}(x) + (1 - p) \Delta_{\mathrm{-log}\frac{1-p}{p}}(x)$, and $\Delta_t(x)=\delta_{\textrm{dirac}}(x-t)$.

\begin{figure}
\centering
\includegraphics[width=0.6\linewidth]{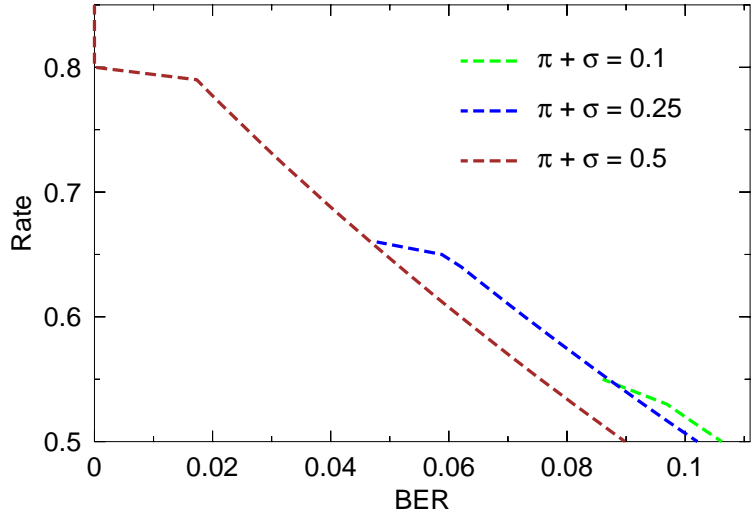}
\caption{Theoretical threshold.}
\label{fig:threshold}
\end{figure}

On Fig.~\ref{fig:threshold} we track the evolution of the threshold for the code with rate one half in~\cite{Elkouss_09}, it can be observed how different values of $\delta$ offer a tradeoff between the range of rates achievable and the efficiency.

In~\cite{Richardson_01a} it is presented a condition for decoding stability:

\begin{equation}
\lambda'(0) \rho'(1) < \frac{1}{e^{-r}}
\label{eq:er}
\end{equation}

\noindent where $e^{-r}$ is defined as:

\begin{equation}
e^{-r} = \int_{\Re} p_{u_0}(x) e^{-x/2} dx
\end{equation}

\noindent operating:

\begin{equation}
\lambda_2 < \frac{1}{(2 \sqrt{p (1 - p)} (1 - \delta) + \pi) \rho'(1)}
\label{eq:stability}
\end{equation}

\noindent which imposes a limitation when choosing a code: it has to be stable for the whole range of rates in which it will be used. A code with $\lambda_2$ close to the stability limit for $R_0$ can become unstable for for high values of $\pi$. 

\section{Simulation Results}
\label{sec-results}

In order to understand the behavior of the protocol described in section~\ref{sec-ratecompir}, we analyze the rate compared to the optimal value.

The family of LDPC codes used in our simulations have been obtained from~\cite{Elkouss_09} and the Tanner graphs have been constructed using a modified Progressive Edge-Growth (PEG) algorithm~\cite{Martinez_09}. This improved PEG construction is based on the original~\cite{Hu_05}, but it also takes into account $\rho(x)$, the check distribution  polynomial. 
We have used a single code of length $n = 200.000$, a reasonable lower bound of the expected length in QKD transmission. Bigger $n$ values would improve the performance of the protocol (by increasing the reconciliation efficiency). The rate is one half, that allows to cover all range of expected BERs. 
Simulations have been done with an LDPC decoder based on belief propagation, with a maximum number of 2000 iterations per simulation. The LDPC decoder has been modified to work with puncturing and shortening, adding two new $\log$-likelihood ratios for the initialization of puncturing, $\gamma_{p} = 0$, and shortening, $\gamma_{s} = \infty$, respectively. The points in the different figures have $p_{\textrm{bit}} < 10^{-6}$.



\begin{figure}
\centering
\includegraphics[width=0.6\linewidth]{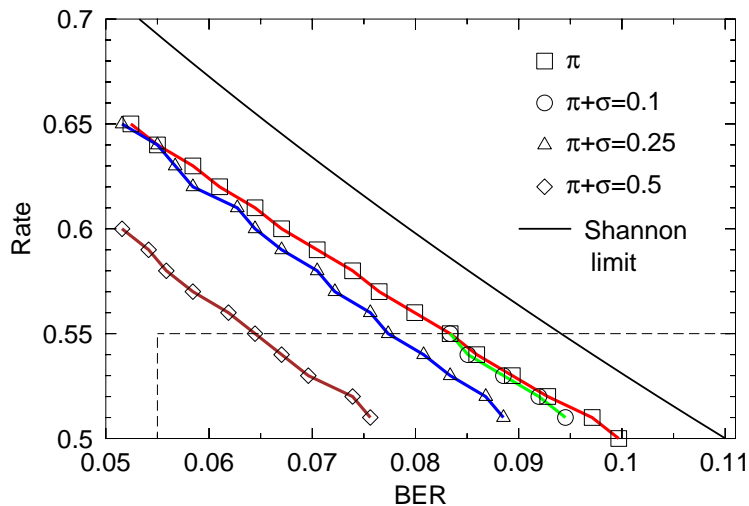}
\caption{Rate achieved over a BSC with $\delta \in \{0.1, 0.25, 0.5\}$.}
\label{fig:punct-50-70}
\end{figure}

\begin{figure}
\centering
\includegraphics[width=0.6\linewidth]{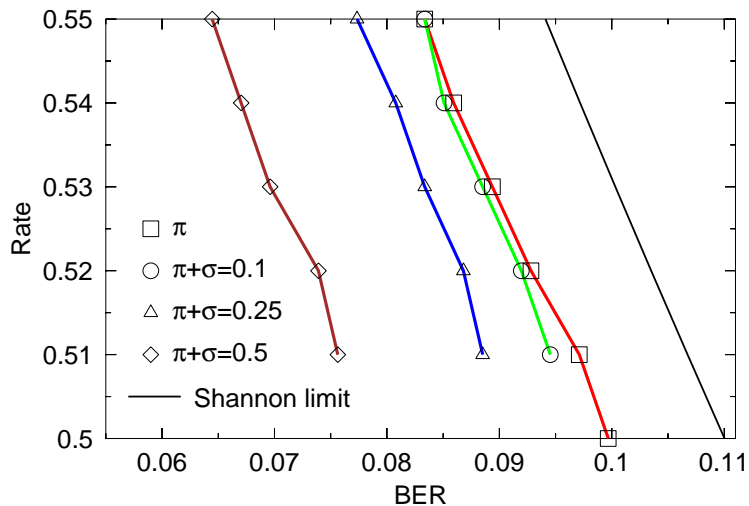}
\caption{Enlarged figure of the portion marked with a dashed line in Fig.~\ref{fig:punct-50-70}.}
\label{fig:punct-50-55}
\end{figure}

In Fig.~\ref{fig:punct-50-70} we present the maximum BER reached over a BSC with the rates going from $R=0.5$ to $0.7$ using different  values of the $\delta$ parameter to regulate the puncturing and shortening. The strong dependence of the rate with parameter $\delta$ is clearly seen. This figure shows the rate achievable for $\delta \in \{0.1, 0.25, 0.5\}$, and it is compared with the rate achieved by the code in the case that it were only punctured and with the Shannon limit. 
These results highlight that, once the reconciliation problem has been characterized and it is known the range of possible error rates, $\delta$ should be chosen as small as possible. If $\delta$ is found to be too big, then it should be considered enlarging the pool with codes that cover different rates. This behaviour can be more clearly seen in the enlarged figure (Fig.~\ref{fig:punct-50-55}) displaying the rate range from $R=0.5$ to $0.55$. The minimum value of $\delta$ that allows to cover the entire interval is $\delta = 0.1$. For this value the decoding performance is similar to~\cite{Elkouss_09}. However, with this protocol we are able to reconcile a continuum of crossover probabilities. For the other values of $\delta$, the performance is worse, however it should be noted that carefully choosing which symbols should be punctured and which ones shortened could improve on these results \cite{Ha_04, Richter_06, Klink_08}.

Looking at 
Table~\ref{tab:efficiency} we can see the effect of the protocol on the efficiency of the reconciliation. When close enough to $R_0$ it is close to one, and for small enough $\delta$ values it remains close to one for the whole set of rates, which is not the case for the higher $\delta$ values as expected by the thresholds found in Fig.~\ref{fig:threshold}. 

\begin{figure}
\centering
\includegraphics[width=0.6\linewidth]{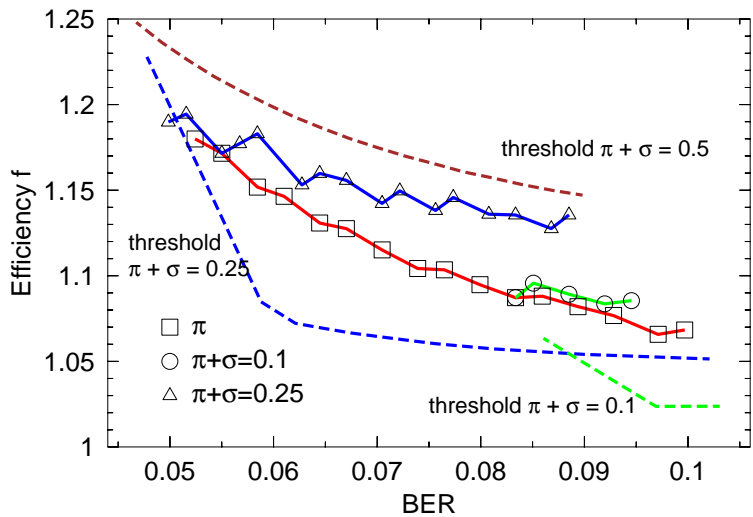}
\caption{Reconciliation efficiency calculated from Eq.~\ref{eq:efficiency}.}
\label{fig:efficiency}
\end{figure}

\begin{table}
\caption{Efficiency calculated from Eq.~\ref{eq:efficiency}.}
\label{tab:efficiency}
\begin{minipage}{1.0\linewidth}
\centering
\begin{tabular}{l l l l l l l}
\hline
\bfseries & \multicolumn{2}{c}{$\delta = 0.1$} & \multicolumn{2}{c}{$\delta = 0.25$} & \multicolumn{2}{c}{$\delta = 0.5$} \\
\multicolumn{1}{c}{$R$\footnote{Rate after puncturing and shortening.}} & \multicolumn{1}{c}{BER\footnote{Maximum bit error rate corrected.}} & \multicolumn{1}{c}{$f$\footnote{Corresponding efficiency for random puncturing and shortening.}} & \multicolumn{1}{c}{BER} & \multicolumn{1}{c}{$f$} & \multicolumn{1}{c}{BER} & \multicolumn{1}{c}{$f$} \\
\hline
0.51 & 0.0945 & 1.0855 & 0.0885 & 1.1356 & 0.0756 & 1.2675 \\
0.52 & 0.092  & 1.0836 & 0.0868 & 1.1276 & 0.0739 & 1.262  \\
0.53 & 0.0885 & 1.0892 & 0.0834 & 1.1355 & 0.0696 & 1.2895 \\
0.54 & 0.0851 & 1.0957 & 0.0808 & 1.136  & 0.067  & 1.2966 \\
0.55 & 0.0834 & 1.0877 & 0.0773 & 1.1457 & 0.0645 & 1.3048 \\
0.56 &        &        & 0.0756 & 1.1382 & 0.0619 & 1.314  \\
0.57 &        &        & 0.0722 & 1.1496 & 0.0584 & 1.3386 \\
0.58 &        &        & 0.0705 & 1.1423 & 0.0559 & 1.3513 \\
0.59 &        &        & 0.067  & 1.1557 & 0.0541 & 1.3659 \\
0.6  &        &        & 0.0645 & 1.1598 & 0.0516 & 1.3651 \\
0.61 &        &        & 0.0627 & 1.1531 \\
0.62 &        &        & 0.0584 & 1.183  \\
0.63 &        &        & 0.0567 & 1.1772 \\
0.64 &        &        & 0.055  & 1.1715 \\
0.65 &        &        & 0.0516 & 1.1945 \\
\hline
\end{tabular}
\end{minipage}
\end{table}

\section{Conclusion}

We have demonstrated how to adapt an LDPC code for rate compatibility. The capability to adapt to different error rates while minimizing the amount of published information is an important feature for QKD key reconciliation. The present protocol alows to reach efficiencies close to one while limiting the information leakage and having the important practical advantage of low interactivity.

Future work will concentrate on the optimization of the puncturing and shortening processes, now done randomly.

\section*{Acknowledgment}
This work has been partially supported by grant UPM/CAM Q061005127.

The authors acknowledge the computer resources and assistance provided by Centro de Supercomputaci\'on y Visualizaci\'on de Madrid (CeSViMa).

\bibliographystyle{IEEEtran}
\bibliography{itw2010}

\vfill

\end{document}